\documentclass[aip,apl,superscriptaddress,amsmath,amssymb,reprint]{revtex4-2}

\usepackage{tikz}
\usepackage{graphicx}
\usepackage{dcolumn}
\usepackage{bm}
\usepackage{float}
\usepackage{svg}
\usepackage{xr}
\usepackage{braket}
\usepackage{bbold}
\usepackage{enumerate}
\usepackage{amsmath}
\usepackage{mathtools}
\usepackage{wrapfig}
\usepackage{amsmath} 
\usepackage{amssymb} 
\usepackage{seqsplit}
\usepackage{xcolor}
\usepackage{braket}
\usepackage{tabularx}
\usepackage{makecell}
\usepackage[export]{adjustbox}
\usepackage[shortlabels]{enumitem}
\usepackage{setspace}
\usepackage{tensor}
\usepackage{hyperref}
\usepackage[capitalise]{cleveref}

\newcommand{\bes} {\begin{subequations}}
\newcommand{\ees} {\end{subequations}}
\newcommand{\beq}{\begin{equation}}
\newcommand{\eeq}{\end{equation}}

\def\theYear{\the\year}
\graphicspath{ {./figures/} }
\hypersetup{pdfpagemode=FullScreen, colorlinks=true, allcolors=blue}

\begin{document}
\title{Simple, accurate lumped-element models of distributed resonators for superconducting quantum circuits}

\author{Elizabeth Kunz}
\email{ekunz@usc.edu}
\affiliation{Center for Quantum Information Science and Technology, University of Southern California, Los Angeles, California 90089}
\affiliation{Department of Physics \& Astronomy, University of Southern California, Los Angeles, California 90089}

\author{Eli M. Levenson-Falk}
\email{elevenso@usc.edu}
\affiliation{Center for Quantum Information Science and Technology, University of Southern California, Los Angeles, California 90089}
\affiliation{Department of Physics \& Astronomy, University of Southern California, Los Angeles, California 90089}
\affiliation{Ming Hsieh Department of Electrical \& Computer Engineering, University of Southern California, Los Angeles, California 90089}

\begin{abstract}
    Superconducting quantum circuit design is reliant on accurately mapping design parameters to a quantum Hamiltonian. Designers typically rely on computationally intensive finite-element electromagnetic simulations. However, effective circuit-level models can in principle capture much of the relevant physics for these systems, reducing the need for finite-element simulations. A barrier to implementing these simpler models has been the prevalence of distributed elements such as coplanar waveguides resonators in device designs. Techniques for modeling these distributed elements have been developed, but may be difficult to scale, reduce intuition, and give inaccurate results under strong coupling. In this work we describe a simple effective circuit-level model that faithfully reproduces the scattering parameters, and subsequently can predict certain Hamiltonian parameters, for a coupled, distributed element within a broader two-port network even up to strong coupling. Our approach does not explicitly require an electromagnetic simulation. Our model, along with other common lumped models used in black box quantization, is publicly available to researchers via an open-source code package, $\texttt{simpleLOMs}$. 
\end{abstract}

\maketitle

A superconducting device's behavior depends intricately on the circuit design \cite{LevensonFalk2025_DesignConcerns}. To predict this behavior designers typically take a lumped-element circuit model and \textit{quantize} it, constructing a quantum circuit Hamiltonian \cite{DevoretVool2017,rasmussen_superconducting_2021}. They then transform charge and flux operators into mode operators, creating an \textit{effective Hamiltonian} of qubits/qudits/oscillators with inter-mode couplings \cite{krantz_quantum_2019}. There are established recipes for circuit quantization \cite{DevoretVool2017, Kerman2020, Rajabzadeh2023_SQcircuit}. However, modern superconducting quantum devices include distributed elements. These may be modes of a 3D cavity \cite{Paik2011}, transmission lines \cite{Blais2004}, or quasi-lumped capacitors and inductors whose size is not negligible compared to the wavelength \cite{Peruzzo2021}. It can be challenging to quantize a circuit containing these elements. To address this challenge, three techniques are common. A lumped oscillator model (LOM) approach treats a selected mode of each distributed element as an LC oscillator, giving a lumped model that can be quantized with standard techniques \cite{DevoretVool2017,you_circuit_2019}. This approach is often used to model transmission line resonators in planar circuits: the resonator is modeled as an effective LC circuit using analytical formulas that assume open or shorted ends, although some approaches consider loading at one end \cite{Minev2021_quasiLOM}. Energy participation ratio (EPR) approaches model the entire device, then calculate the fraction of each mode's energies that are stored in each nonlinear element to directly generate an effective Hamiltonian \cite{Minev2021_EPR}. Blending EPR and LOM is black box quantization (BBQ), which separates out the nonlinear elements and constructs a simplified lumped-element model for the rest of the circuit, attempting to match its impedance \cite{Nigg2012_BBQ}. 

Contrary to popular perception, none of these techniques require a full electromagnetic (EM) simulation of the device geometry. Instead, one can calculate the classical response of an effective circuit model comprising both lumped and distributed elements using rapid numerical techniques. This approach has been applied to, e.g., perform EPR analysis on circuit models \cite{Minev2021_EPR, Zaccaria2025}, to rapidly predict the resonant frequency and linewidth of a tightly-wound transmission line resonator \cite{Larry_thesis}, and to directly calculate effective Hamiltonian parameters based on impedance \cite{solgun_simple_2019}. However, in many circumstances it is desirable to have a lumped-element circuit model, as this permits analytical expressions for many effective Hamiltonian quantities and can give clearer intuition as to how different parameters affect device behavior.

In this paper we present a rapid, straightforward procedure for creating simple, accurate lumped-element models of distributed resonators in quantum circuits. In contrast to standard LOM, we model the distributed resonator in full awareness of lumped coupling elements attached to it. In contrast to common BBQ approaches, we consider only a single distributed element at one time and replace \textit{only} the distributed element with an effective lumped model, keeping lumped coupling elements unchanged. This allows an effective model to be created self-consistently while not changing other elements when distributed elements are added. Our procedure maintains accuracy even under heavy loading, beating out both traditional LOM and BBQ using single-pole Foster synthesis. We share an open-source code package that allows users to implement the procedure easily \cite{simpleLOMs}. Our results provide a straightforward way to implement lumped models and generalize easily to larger circuits.

\begin{figure}
    \includegraphics[width=3.3in]{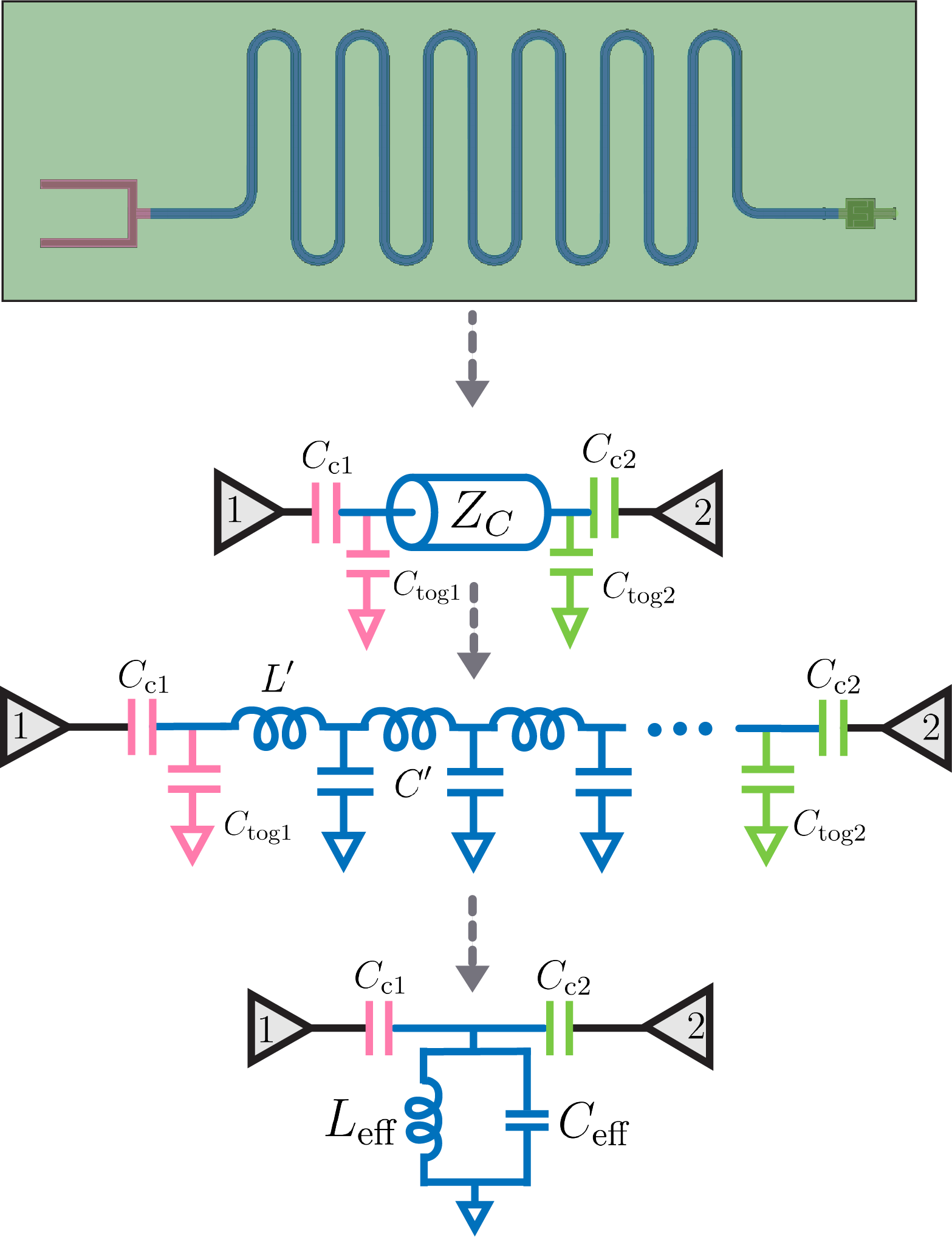}
    \caption{\label{fig:diagram} A distributed element such as a coplanar waveguide can be represented by an abstract transmission line with characteristic impedance $Z_C$. It is capacitively coupled at the ends to arbitrary loads by $\{C_{\text{c1}}, C_{\text{c2}}\}$ and to ground by $\{C_{\text{tog1}}, C_{\text{tog2}}\}$. Exactly equivalent to this transmission line is an infinite chain of infinitesimal inductors in series and capacitors to ground. The first mode of this chain can be approximated by a single resonant circuit $LC$ with effective inductance $L_{\text{eff}}$ and capacitance $C_{\text{eff}}$. }
\end{figure}

We consider the setup in \cref{fig:diagram}. A coplanar waveguide (CPW) is terminated at either end with capacitors. This may be, e.g., a readout resonator coupled to a qubit and a measurement feedline or Purcell filter \cite{Reed2010_PurcellFilter, Jeffrey2014_PurcellFilter}. We represent it as an abstract transmission line with impedance $Z_C$, length $L$, phase velocity $v$, and coupling capacitors $C_\mathrm{c1}, \ C_\mathrm{c2}$ to ports 1 and 2. Each capacitor also contributes ground capacitance $C_\mathrm{tog1,2}$. Such a transmission line is equivalent to an infinite chain of infinitesimal inductors in series and capacitors to ground. However, for circuit quantization to be computationally tractable we need to limit the number of circuit nodes, i.e., to use the smallest equivalent circuit that gives the required accuracy. We focus on the simplest and most common case, where the transmission line is approximated as a single parallel $LC$.

A typical approach is to treat the transmission line as if it were open-circuit or short-circuit on both ends. In this model, the $n$th mode of the transmission line can be equivalently represented as a parallel LC resonator with values
\begin{equation}
    \label{eq:analyticalLC}
    C_\mathrm{eff}^\mathrm{Analytic} = \frac{n\pi}{mZ_c\omega_r}~~~L_\mathrm{eff}^\mathrm{Analytic}= \frac{nmZ_c}{\pi \omega_r}
\end{equation}
where $Z_c$ is the characteristic impedance, $\omega_r$ is the resonant frequency of the $n$th mode, and $m=2$ if both ends are shorted or open, as is the case with a $\lambda/2$ resonator, or $m=4$ if one end is shorted and the other is open, as is the case with a $\lambda/4$ resonator \cite{Schuster_thesis, Pozar}. \footnote{Note that this expression is derived assuming that the voltage coordinate of the equivalent LC circuit is the same as the voltage at the voltage antinode of the transmission line. This is appropriate when one wants to model capacitive couplings. A different expression is found if one tries to find an equivalent inductance using the current coordinate at the current antinode, as is needed for inductive coupling. We will use the values in \cref{eq:analyticalLC} for this paper as we are only modeling capacitive couplings; our approach should generalize to inductive coupling.}

\cref{eq:analyticalLC} was derived assuming that the voltage and current vary sinusoidally along the transmission line. This approach can lead to inaccurate results when one or both ends of the transmission line are strongly loaded---that is, when they are terminated with capacitances or inductances that are neither much smaller nor much larger than the effective transmission line capacitance or inductance---as this causes distortions of the mode's spatial structure.  A more accurate quantization can be achieved by first finding the eigenmodes of the \textit{loaded} transmission line, then using EPR analysis to quantize effective mode charge and flux variables \cite{Minev2021_EPR} (as opposed to circuit node charge and flux variables). This has the advantage and limitation of bypassing the lumped element equivalent model entirely, making the problem simpler but also making the result less generalizable. Another approach is to make phenomenological adjustments to \cref{eq:analyticalLC}. For example, in the SQuADDS database a transmission line terminated with coupling capacitors on either end is first modeled by terminating with a capacitor and a short/open, using that resonant frequency with \cref{eq:analyticalLC}, then adding the other capacitor back in \cite{Shanto2024_SQUADDS}. This provides a more accurate lumped-element model but does is not self-consistent, as the model depends on which side of the resonator one decides to short/open.

The final approach is black box quantization, and our technique may be seen as a type of BBQ. In traditional BBQ, the nonlinear inductances of Josephson elements are isolated from the rest of the circuit (including the linear Josephson inductance). This circuit is turned into a ``black box'' with impedance $Z(\omega)$. The black box is in turn modeled as a lumped circuit, typically using Foster synthesis: a parallel RLC circuit is created that approximately matches the impedance near a pole or zero of the impedance or admittance (i.e., near a resonance), then cascaded with other RLC circuits for a finite number of other poles. The simple model is to set an effective capacitance 
\begin{equation*}
    \left. C_\mathrm{eff}^\mathrm{Foster} = \frac{1}{2}\frac{d}{d\omega}\mathrm{Im}\{Y(\omega)\}\right|_{\omega_n} ~~~ L_\mathrm{eff}^\mathrm{Foster} = \frac{1}{\omega_n^2 C_\mathrm{eff}^\mathrm{Foster}}
\end{equation*}

where $Y(\omega)$ is the complex admittance and $\omega_n$ is the frequency of the $n$th pole. A more sophisticated approach, Brune synthesis, uses a more complex circuit geometry but can achieve a better fit over a broader frequency range \cite{Solgun2014_BBQ_with_Brune_synthesis}. However, using these approaches to model a network with multiple ports of connection can require a complicated circuit \cite{Nigg2012_BBQ}. There has been little study of how accurate these multi-port models are when a new load is placed on a port, e.g., a resonator or qubit is added to the network. While a new effective model can always be computed, it is often desirable to add on to an existing model without needing to reconstruct it from scratch. Likewise, the elimination of coupling elements in the effective circuit can make it less straightforward to interpret some results.

We attempt to use the simplest possible circuit to match the behavior of the distributed elements from all ports while leaving  lumped elements unchanged. Again, we consider the setup in \cref{fig:diagram}: a transmission line resonator terminated with coupling capacitors both ends. Unlike standard BBQ approaches where the whole circuit including coupling capacitors would be replaced or modified, we replace \textit{only} the transmission line with a single lumped LC circuit, leaving the coupling capacitors unchanged. The capacitances to ground add in parallel with $C_\mathrm{eff}$ and so we absorb them into $C_\mathrm{eff}$. Unlike the analytical approach taken to derive \cref{eq:analyticalLC}, we find $C_\mathrm{eff}$ and $L_\mathrm{eff}$ taking into account the coupling capacitors. We do this by taking the transmission line with capacitors, adding ports to either end, and simulating scattering ($S$) parameters using the \texttt{scikit-rf} package \cite{Arsenovic2022_Scikit_RF}. We then make an $LC$ circuit with the same coupling capacitances, and vary $L_\mathrm{eff}$ and $C_\mathrm{eff}$ to fit the transmission line S parameters as closely as possible within a range of $f_0 \pm 5\kappa$, where $f_0$ is the resonant frequency and $\kappa$ is the linewidth.  We choose the fit frequency range to be large enough to capture the distributed behavior of the CPW but small enough to avoid fitting to only the background impedance. The fit is performed using \cref{eq:analyticalLC} as an initial seed guess, then using least squares regression on the real parts of $S_{11}$ and $S_{22}$ simultaneously \footnote{We neglect the imaginary parts as these are analytical functions and so are fully determined by their real parts; likewise, $S_{12}$ and $S_{21}$ are fully determined by $S_{11}$ and $S_{22}$.}.

\begin{figure*}
    \includegraphics[width=7in]{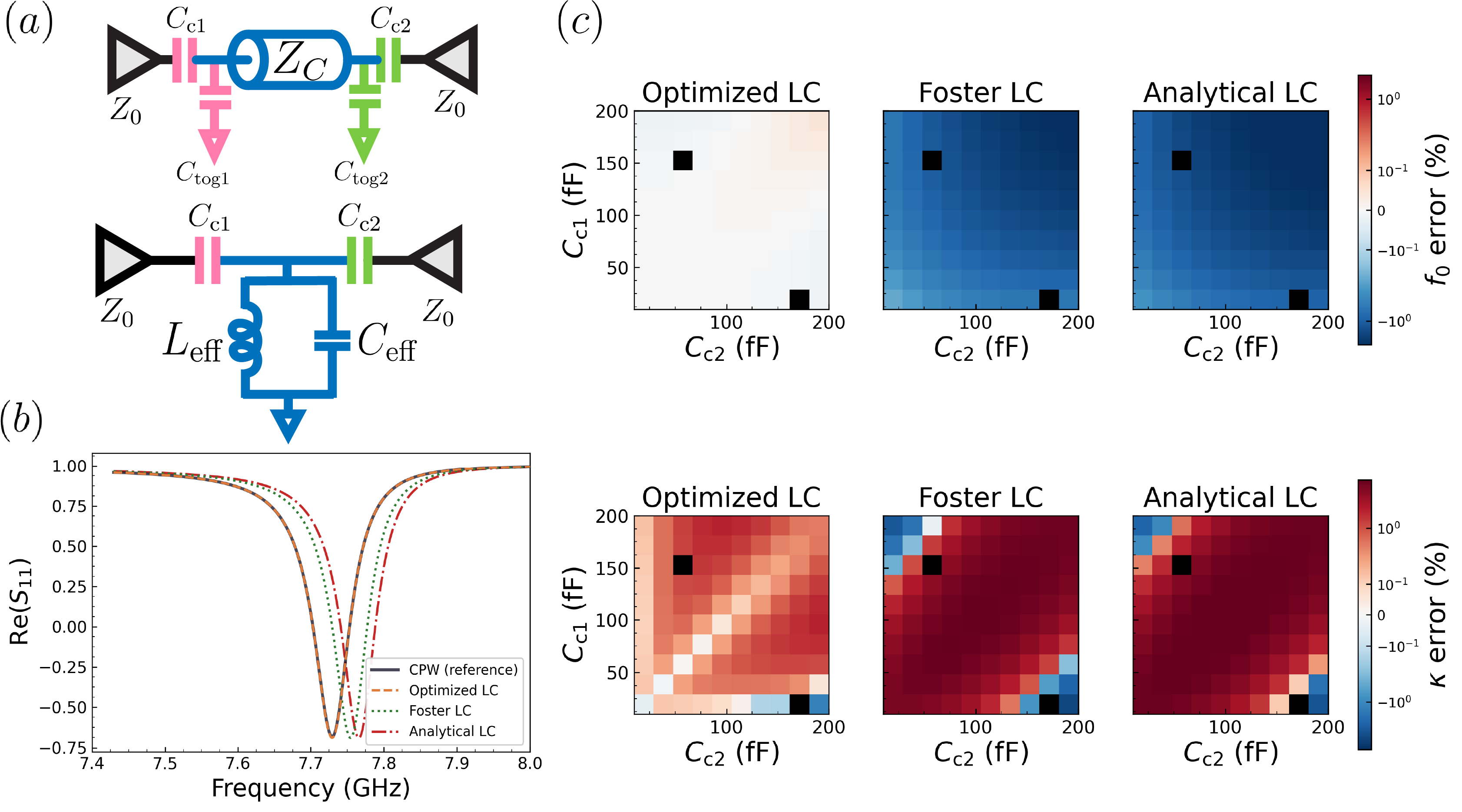}
    \caption{\label{fig:Scomparison}  $(a)$ Circuit diagrams of a distributed system with characteristic impedance $Z_C$, and its corresponding effective lumped circuit model, with effective capacitance and inductance $\{C_{\text{eff}}, L_{\text{eff}} \}$. $(b)$ $\text{Re}(S_{11})$ transmission spectrum of the Optimized, Analytical, and Foster models, with the distributed CPW element included as a reference. $(c)$ Heat maps (log scale) of the percent error of each method in predicting the distributed circuit shifted frequency $f_0$ and linewidth $\kappa$, as measured via a circle fit performed on $S_{11}$ in the complex plane. Blacked-out points indicate points where there were errors that prevented fitting of the resonance of at least one model.}
\end{figure*}

We show a comparison of our technique with the analytical LOM and Foster synthesis approaches in \cref{fig:Scomparison}. We calculate the port scattering parameters $S_{11}$ and $S_{22}$ for the transmission line model and the equivalent models. Note that for traditional one-port Foster synthesis, the circuit generated depends on whether one calculates admittance from port 1 or port 2; here we choose to use port 1. Our simple, coupling-capacitor-aware fitting approach outperforms the other techniques, matching the response more closely as shown in \cref{fig:Scomparison}(b). We quantify this match by using the circle fit method to fit the frequency $f_0$ and linewidth $\kappa$ of the resulting resonance \cite{khalil_analysis_2012,probst_efficient_2015}. \cref{fig:Scomparison}(c-d) shows the error in frequency and linewidth for all 3 approaches as a function of coupling capacitances compared to the values obtained for a CPW $7000 \text{ um}$ long, characteristic impedance of $Z_0 = 45.92 \ \Omega$, and bare resonant frequency of $f_0 = 8.58 \ \text{GHz}$, with ground capacitances fixed at 20 fF. We attribute the difference in model performance to the fact that our model takes into account the loading on both ends of the transmission line by fitting the system with the coupling capacitors explicitly included.

\begin{figure*}
    \includegraphics[width=7in]{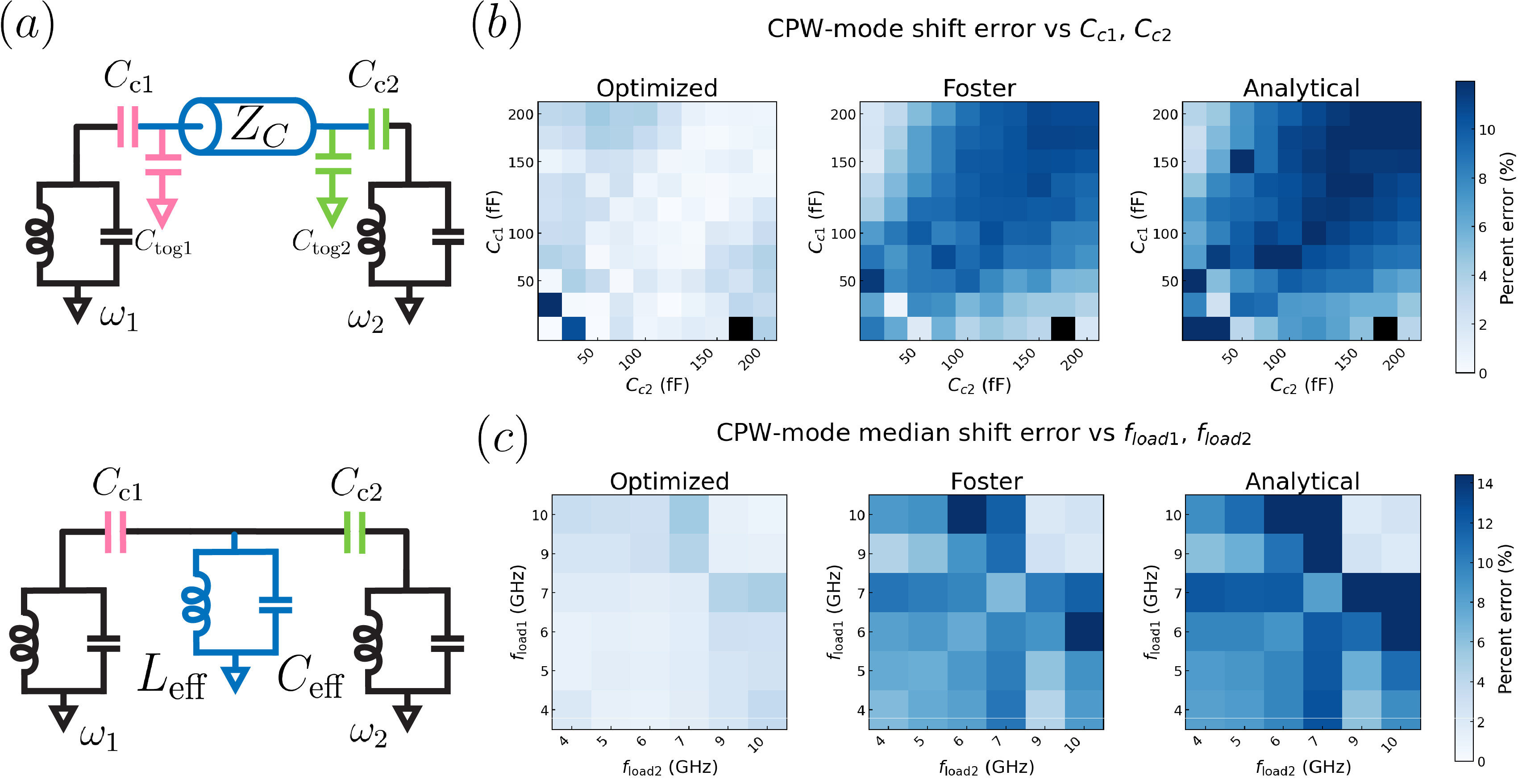}
    \caption{\label{fig:shiftComparison}  $(a)$ Circuit diagrams for the loaded system. While the central resonators and coupling capacitors have not changed, both the distributed and effective circuits are coupled to resonant loads on either end, represented here by parallel $LC$ circuits. $(b)$ Percent error of frequency shift of the effective $LC$ resonance compared to the CPW resonance as a function of coupling capacitance. The off-resonant loads were fixed at $5 \text{ GHz}$ and  $6 \text{ GHz}$, respectively. $(c)$ For each load pairing we find the median shift error across all combinations of capacitances, then plot it as a function of load pairing. We neglect shifts lower than $1 \text{ MHz}$ to avoid numerical errors, as this approaches the difference between points in our sweep.  }
\end{figure*}

A more challenging test is whether our model produces appropriate couplings to far-off-resonant loads. To check this, we simulate a situation where lumped-element resonators are coupled on either end of the transmission line as shown in \cref{fig:shiftComparison}(a). The resonant frequencies of these loading resonators and the central CPW resonator all shift due to the coupling. This shift is a proxy for the coupling strength $g$ between two resonators: when the difference between their bare resonant frequencies $\omega_1,\ \omega_2$ is much larger than the coupling, the shift $\chi_L$ is approximately
\begin{equation}
    \chi_L \approx g^2 \left(\frac{1}{\omega_1-\omega_2} - \frac{1}{\omega_1+\omega_2}\right)
\end{equation}
where this expression was derived in second order perturbation theory \cite{Shanto2024_SQUADDS}.

 A comparison between this optimized model and traditional lumped models as a function of coupling capacitances and load resonator frequencies is shown in  \cref{fig:shiftComparison}. For the central resonator, we use the same dimensions as in \cref{fig:Scomparison}. We note that the higher end of the capacitance sweep (200 fF) is near the limit of what is common in experiments, as typical superconducting devices have a total capacitance budget on the order of hundreds of $fF$ \cite{Koch2007, LevensonFalk2025_DesignConcerns}. We use the same effective $LC$ models that were used in \cref{fig:Scomparison}, i.e., derived by fitting the response in the presence of capacitively coupled ports, not in the presence of resonant loads. We again compare the accuracy to the CPW model of different models in predicting the shifted resonant frequency under the influence of the loads. Specifically we take the shift $\Delta f$ of the resonant frequency when the ports are replaced with resonant loads, and define error as $\epsilon = (\Delta f_\mathrm{LC} - \Delta f_\mathrm{CPW})/\Delta f_\mathrm{CPW}$. We see that in virtually all cases, our approach outperforms the Foster and analytical models for the effective $LC$, more accurately predicting the shift in CPW frequency.

\begin{figure}[h!]
    \includegraphics[width=3.3in]{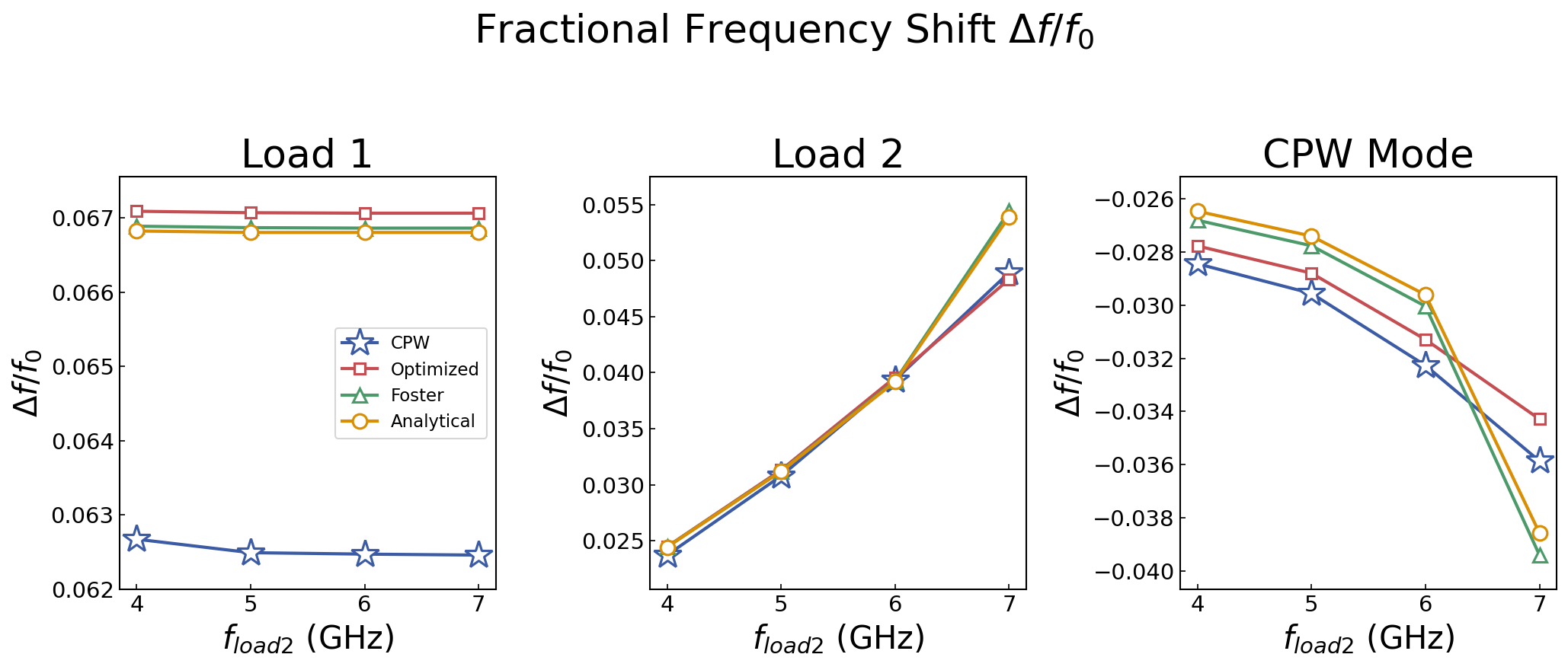}
    \caption{\label{fig:fractional_shiftComparison} Fractional shift $\Delta f / f_0$ for each mode as a function of the bare frequency of the second load with a fixed pair of coupling capacitors at $C_{\text{c1}} = \text{157.8 fF}$ and $C_{\text{c2}} = \text{52.2 fF} $. For off-resonant loads such as $5 \text{ GHz}$, all models perform approximately equally well, while loads closer to the CPW resonant frequency are better improved by implementing the Optimized method. As shown earlier, the CPW mode is always better matched by the Optimized method.}
\end{figure}

 A caveat is that we do not always see more accurate prediction of the shifts of the load resonators themselves, as shown in \cref{fig:fractional_shiftComparison}. Different approaches perform best depending on the load resonator frequencies, although they are generally quite similar to each other and distinct from the true CPW model. We attribute this to the fact that such off-resonant shifts are relatively \textit{insensitive} to small changes in $C_{\text{eff}}$ and $L_{\text{eff}}$, and the different lumped models give relatively similar values for these parameters. This effect can be seen in the Load 1 and 2 modes, where the fractional frequency shift is not strongly affected by the choice of model until $f_{\text{load2}}$ approaches the CPW mode frequency. Meanwhile the third mode associated with the distributed CPW is affected across all choices of $f_{\text{load2}}$, showing its greater sensitivity to model choice---in this case our approach performs best.

\color{black}

The utility of our method would be limited if the numerically-solved port parameters of the transmission line did not accurately represent reality. Fortunately, recent demonstrations have shown that there can be an excellent match even in the case where the transmission line meanders, coupling to itself at different points, as is common in many device geometries \cite{Larry_thesis}. As a reconfirmation we simulate the simple case where the transmission line is straight and coupling capacitors are lumped elements. We perform a simulation of the eigenmodes of the transmission line with virtual lumped capacitors and ports at the ends using the \texttt{ANSYS HFSS} (High Frequency Structure Simulator) driven modal simulator. We simulate a straight CPW with zero thickness, coupled on both ends to $50 \Omega$ loads by lumped capacitors, with additional 10 fF capacitance to ground on either end. We test the capacitances $C_c = \{ 20 \text{ fF}, 50 \text{ fF}, 100 \text{ fF} \}$. We then compare the results of this driven modal simulation to the optimized lumped model constructed in $\texttt{scikit-rf}$. We find that the optimized model is accurate to within below 0.5\% of the shifted frequency and  below 3\%  of the linewidth in all tested cases.

In conclusion, we have demonstrated a technique that replaces a distributed element resonator with a single LC circuit, without changing the values of nearby circuit parameters. The technique is more accurate than traditional lumped model approaches when heavily loaded because it takes into account the loading elements when constructing the LC. The technique is also more flexible than traditional BBQ methods that replace all linear circuit elements with a single effective circuit, since it can be applied modularly across a circuit. The method does not require the user to define a single reference port, and generates the same effective circuit whether measured in $S_{11}$ or $S_{22}$. We have deployed our technique in a simple code package, titled $\texttt{simpleLOMs}$ which can be used by researchers to make their designs \cite{simpleLOMs}. The code is modular so that any 2-port microwave network can be substituted for the transmission line used in this work. Likewise, a more sophisticated model with multiple LC resonators can be deployed to handle higher modes, as is typically done in Foster synthesis. We include tutorials and examples in the documentation, including showing how our approach can be cascaded across a network of many distributed-element resonators \cite{noauthor_simpleloms_nodate}.

Extensions to more complicated geometries are relatively straightforward. For instance, a circuit with coupling midway along a transmission line can be modeled as a transmission line on either side of the coupling element. Likewise, a network with many ports can be modeled using a series of parallel LC resonators with the L and C values fit to the scattering responses from all ports in the presence of the couplings. We will explore these situations in future work. Future work could also explore how best to implement new techniques for modeling time-dependent drives on distributed circuits \cite{lu_systematic_2026}. Our results show that simple lumped models are a viable approach for accurate circuit-level simulation of a superconducting device.

We gratefully acknowledge Sadman Shanto, Larry Chen, Zlatko Minev, and Saikat Das for useful discussions. Funding was provided by the Army Research Office under grant W911NF-25-1-0255 and by the National Science Foundation under grant 2612093.

\bibliography{references}

\end{document}